\documentclass[twocolumn,showpacs,preprintnumbers,amsmath,amssymb]{revtex4}
\usepackage{graphicx}

\begin{document}

\def\bea{\begin{eqnarray}}
\def\eea{\end{eqnarray}}
\def\a{\alpha}
\def\p{\partial} 
\def\nn{\nonumber}
\def\r{\rho}
\def\xb{\bar{x}}
\def\vb{\bar{v}}
\def\fb{\bar{f}}
\def\lab{\bar{\lambda}}
\def\la{\langle}
\def\ra{\rangle}
\def\f{\frac}
\def\o{\omega}
\def\P{\mathcal{P}}
\def\d{\delta}

\title{
Spin Precession and Avalanches
}

\author{J. M. Deutsch$^1$ and A. Berger$^2$}
\affiliation{
$^1$Department of Physics, University of California, Santa Cruz, California 95064}

\affiliation{
$^2$San Jose Research Center, Hitachi Global Storage Technologies, San Jose, California 95120
}

\date{\today}

\begin{abstract}
In many magnetic materials, spin dynamics at short times are dominated 
by precessional motion as damping is relatively small. In the limit
of no damping and no thermal noise, we show that for a large enough
initial instability, an avalanche can transition to an ergodic phase
where the state is equivalent to one at finite temperature, often above
that for ferromagnetic ordering. This dynamical nucleation phenomenon is
analyzed theoretically. For small finite damping the high temperature
growth front becomes spread out over a large region. The implications
for real materials are discussed.
\end{abstract}


\maketitle

Ferromagnetic systems that 
are subject to slowly changing external magnetic fields very commonly show 
avalanche-like responses ~\cite{Barkhausen}.  
This leads to
hysteresis, as avalanches occur over a very fast timescale resulting in
irreversibility and entropy production.

A large amount of experimental and theoretical work has been
devoted to understanding aspects of this behavior, 
such as Barkhausen noise~\cite{Barkhausen} which demonstrates
that there is often a large degree of reproducibility
in the mesoscopic dynamics on repeated cycling of the 
field and also interesting critical properties~\cite{Sethna,SethnaReview}.
With advances in experimental techniques, direct tests of the reproducibility
of magnetic memory have been undertaken recently which highlighted the
prominent role of sample disorder~\cite{Sorensen}.

The theoretical treatments to date have largely relied on simplified
models such as the Ising model to understand these complex systems. 
The dynamics of such models have been purely relaxational, the extreme
limit of large damping, whereby a spin
is flipped if the energy of the system is decreased by doing so,
the excess energy being transferred out of spin degrees of freedom, to
for example, phonons. 
These models have had a great deal of success in describing many
features of disorder ferromagnets, showing
fascinating properties, for instance ``Return Point Memory" (RPM)~\cite{Sethna,SethnaReview}. 

However real
magnets are typically dominated by precessional effects
on short enough time scales. The
Landau-Lifshitz-Gilbert (LLG) equation~\cite{LLGref}, 
contains a precessional term and a dissipative one
\begin{equation}
\frac{d{\bf s}}{dt} = -{\bf s} \times ({\bf B} 
-\gamma {\bf s} \times {\bf B}),
\label{eq:LLG}
\end{equation}
where ${\bf s}$ is a microscopic magnetic moment, ${\bf B}$ is the
local effective field, and $\gamma$ is a damping coefficient.
$\gamma$ measures the relative importance of damping to precession.
It ranges from about .01 to 1 in real materials~\cite{GammaValues}. 
Therefore in many magnetic
materials, there should be an interesting short time regime where 
it makes sense to regard the damping as a perturbation.
In fact, we will show 
that there is a short time scale in which the magnetic response to the
applied field change is strongly influenced by the finite
level of damping in real materials. 
Furthermore, even macroscopic and long-time properties, such as 
hysteresis loops are influenced, by the level of damping and other
materials properties, that we shall explain in more detail below. 

Throughout this paper, we only consider the case of zero thermal noise.
This is because we will see that {\em effective} finite temperature 
behavior is found even in this case and we want to carefully
separate out these two effects.

If we adiabatically lower the external field, 
taken to be in the $z$ direction, 
at some point
the system will go unstable and have an avalanche. This will
involve the 
nonlinear and possibly 
chaotic motion of its spins
that interact through ferromagnetic and dipolar interactions.

We will first analyze what happens during the avalanche when we set
$\gamma$ to zero.  This will describe the dynamics of the system for short
time scales. In this case, the dynamics conserve energy. If the dynamics
are sufficiently nonlinear, we might expect the system to be ergodic,
which means that its equilibrium behavior is well described by the
microcanonical ensemble. In this case, the avalanched state is not static
but is one of a system at a finite temperature, and we will see that this
temperature can be quite high, even above the ferromagnetic transition
temperature.  However there are two reasons why ergodicity may break down.
The first being if the Hamiltonian has strict rotational symmetry about an
axis leading to conservation of angular momentum~\cite{MaMazenko} in that direction~\cite{footnote:invariant}.

The second reason why ergodicity may be broken is more subtle and to
our knowledge, this is the first time a scenario of this kind has been
proposed. After the system initiates an avalanche, energy is transmitted
into neighboring spins, some of which will be in the form of spin waves.
This will propagate energy away from the avalanche region which will 
decrease the temperature of the avalanched spins, implying that
as time progresses, the avalanched
region becomes cooler. So if neighboring
spins are not recruited, the avalanche will be extinguished. In this
sense, the spin waves act as a damping term even if there is no damping
in the LLG equation. For long times, still assuming $\gamma = 0$, the
energy will be distributed in all the degrees of freedom of
the entire system, which means in the limit of infinite system size, the temperature
of the system will have dropped back down to zero. One is then left
with a system that has produced only a sub-system-size avalanche and has got trapped in 
another local minimum. 

Therefore we propose that an avalanche that is initially large enough,
will propagate through the whole system causing it to go into a state of
statistical mechanical equilibrium, often at a high temperature (for $\gamma = 0$).  
If the initial avalanche is small, the avalanche will usually die out instead,
leading to only a finite number of spins changing the sign of $s_z$.

We now turn to two dimensional numerical experiments to support these claims 
and study the case of finite damping.
Most real experiments on avalanche dynamics have been effectively two dimensional~\cite{SethnaReview}.
Dipolar forces were not included as they complicate the analysis by adding
an additional parameter. Their effects will be the subject of future work.

We consider a Hamiltonian that couples nearest neighbor spins on
a two dimensional square lattice and contains an anisotropy term where the orientation
of the easy axis is randomized slightly about the z-axis and there is disorder
in the ferromagnetic coupling. 
\begin{equation}
\mathcal{H} = -  \sum_{<i,j>}  J_{ij} {\bf s}_i \cdot {\bf s}_j  
-\alpha \sum_{i} ({\bf s}_{i} \cdot {\bf \hat n}_i)^2
-B_{ext} \sum_i s_{i,z}
\end{equation}
where the $J_{ij}$'s are the ferromagnetic coupling constants drawn
from a uniform distribution with nonzero positive mean.  $\alpha$ is a measure
of the anisotropy. We choose the ${\hat n}_i$ to be random but biased towards the
$z$-axis (out of plane). See ref. \cite{DeutschMai} for details.
This system is placed in an external field $B_{ext}$. 

The system was started at high field and the field was lowered adiabatically
by evolving the system at fixed $B_{ext}$ until, to a high accuracy, there was no further change in
spin variables.
To obtain convergence the damping
was made finite, $\gamma = 1$. After this the field
was lowered again. An avalanche was defined to occur when the maximum
$s_z$ among all the spins changed by a finite amount $\Delta \equiv 1$. 
At that point, a successive approximation
scheme was initiated to find the precise field at which the transition takes
place, further evolution can then proceed using the same procedure. 
When an avalanche of desired size was detected, 
the system was restarted with the same external field and pre-avalanche
configuration but now with a different value of damping and the evolution of the system
was recorded.

\begin{figure}[t]
\includegraphics[width=\hsize]{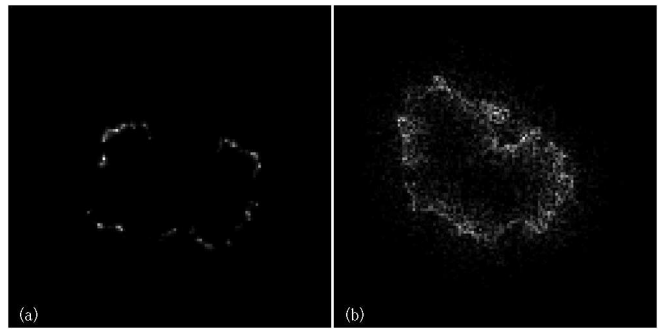}
\caption{
Snapshots of systems with different damping $\gamma$ starting with
the same configuration, during an avalanche. (a) For $\gamma = .8$ and
(b) For $\gamma = 0.01.$ The intensity represents the local spin motion.
}
\label{fig:snapshots}
\end{figure}

A common scenario is to find that the whole system will avalanche
for sufficiently low damping, but will have a sub-system-size
avalanche when the damping is above some critical threshold,
that depends on the precise configuration right before the avalanche.
This is demonstrated for a $128\times 128$ spin system in Fig. \ref{fig:snapshots}.  
At $\gamma = 0.9$, the post-avalanche
cluster of avalanched spins has an approximate diameter of $8$ lattice
spacings. However at $\gamma = 0.8$, the whole system avalanches.
Fig. \ref{fig:snapshots} shows greyscale images of the motion
of the system for $\gamma = 0.8$ (a), and $\gamma = 0.01$ (b),
during their avalanches.
The intensity is proportional to $|d{\bf s}/dt|$. 
In both cases, the
system was started in the same configuration right before
the avalanche. The motion of the system is confined
to the cluster's surface for $\gamma = 0.8$ but is spread out
for $\gamma = 0.01$ in a ring-like structure, 
in which the magnetic system has an elevated effective 
temperature.  In the case of no damping, 
many islands in front of the avalanche's main boundary form and
this elevated temperature range 
encompasses the entire avalanche region. After 
completion of the system-size avalanche, 
the entire system continues to move indefinitely in an ergodic phase.

From these observations, one expects the hysteresis loop to change as one
varies the damping in these simulations. This has been verified directly.

For $\gamma = 0$, the highly chaotic phase is energy conserving and if
there are no other invariants of motion, then the system should be well
described by a microcanonical ensemble. For a large system, this ensemble is well
known to be equivalent to the canonical ensemble (at finite temperature).
We verified by direct simulation of $32 \times 32$ size
systems that this was the case by measuring static and dynamic correlation
functions, for example, $\langle {\bf s}_i(0) \cdot {\bf s}_j(t)\rangle$.

The finite temperature of the avalanche is often higher than
the transition temperature. It
is high because the pre-avalanche metastable configuration has an energy
much higher than the minimum $T=0$ state. When the system gets out of its
trapped static configuration it therefore has a lot of excess energy.
By considering the Ising model with low disorder and estimating the
critical external field for avalanches to take place, we find that
the system can have an energy close to zero, which is consistent
with our numerical results of a high post-avalanche
temperature~\cite{footnote:negativeT}

We next turn to the mechanism by which this ergodic region spreads.
As mentioned above, the effective temperature of the ergodic phase
is quite high, with large amplitude motion over a short time-scale.
The pre-avalanche configuration is static and when these two regions
are connected together, there will be energy transfer between them.
One would expect that over a large scale, Fourier's law should
hold, so that that the temperature in the ergodic region will heat
up the metastable region. Given thermal energy, the metastable region
now has the opportunity to tunnel into the stable phase. 

We will now construct a simple one dimensional model that attempts to
capture the above physics. We use a variable $\phi_i$ to
denote if site $i$ is part of an ergodic region, $\phi_i =1$,
or metastable region, $\phi_i = 0$. There is a temperature
field, which starts off being zero in the metastable region
and a non-zero constant $T_0$ for the ergodic sites that
seed the avalanche. This temperature corresponds to the
energy released per spin when it becomes part of the 
ergodic region. The equation
\begin{equation}
\label{eq:DiffSource}
{\partial T\over \partial t} = D \nabla^2 T +  T_0 {\partial \phi_i\over dt} -
\nu T
\end{equation}
describes thermal diffusion with diffusion coefficient $D$, 
but adds a source term when the region
becomes ergodic. In this case, $\nabla^2$ is a discretized second 
derivative, $\propto T_{i+1}-2T_i+T_{i-1}$. We have also included
for the sake of generality, a last term, $\nu T$, which
is related to the damping in the system. This gives a time scale
for the temperature to die out. For example in fig. \ref{fig:snapshots}(a),
large $\nu$ corresponds to high temperature only on the surface
of the cluster, whereas for low $\nu$, \ref{fig:snapshots}(b), it
persists over a fairly thick surface layer. However the case of $\nu =0$
(i.e. no damping) will be the focus of study below.

The probability per unit time
that $\phi_i$ will go from $0$ to $1$ is $r(t)$, which
we can take for example to be of the Arrhenius form $r_A = \nu_0 \exp(-1/T_i)$,
where $T_i$ is the (dimensionless) temperature on site i.

Together these define a simple model for the disappearance of the
metastable region. We are now in a position to analyze under
what circumstances an avalanche propagates and when it dies out.
We find that for sufficiently large $T_0$ and initial width $w_0$,
of the ergodic seeded region, the avalanche propagates indefinitely, 
but dies out if these two quantities are too small.

As the avalanche propagates, the temperature at the surface will
be  $T_0$ implying that the temperature in the interior 
will be almost constant with the same value. As the temperature
diffuses to sites with $\phi = 0$ , there is a finite probability of
them tunneling to $\phi = 1$.

\begin{figure}[t]
\begin{center}
\includegraphics[width=0.7\hsize]{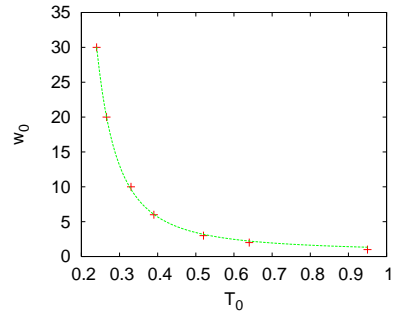}
\caption{
(Color Online)
The boundary between avalanche propagation and extinction for the
one dimensional model discussed in the text. The ``+" symbols are
the numerical values and the solid line is an analytical fit to the data.
}
\label{fig:w_vs_T}
\end{center}
\end{figure}

To estimate the boundary as a function $T_0$ and width $w$, we
assume that for small $T$ that $r(T)$ is a rapidly increasing
function of $T$. If the initial ergodic region is a top hat
function at temperature $T_0$, then for large $w$, there is
a long time when the temperature field  next to the boundary
will approach $T_0/2$. This time $\tau$ will scale as $w^2$. Therefore
the probability $p$ that a site next to the boundary will change
to $\phi=1$ should equal $r(T_0/2)w^2$~\cite{footnote:scaling} for $p << 1$. If it does not 
succeed in tunneling during this time $\tau$, the avalanche
will die out, otherwise it will continue to propagate. As the
avalanche spreads, $w$ increases meaning that $p$ increases so that
it is easier to seed new sites. Therefore we expect the requirement
for an avalanche to propagate is that 
\begin{equation}
w_0 = {c\over r^{1/2}(T_0/2)}
\end{equation}
where $c$ is a constant. We have checked this numerically for the case
of Arrhenius tunneling function $r = r_A$, as mentioned above. The results are shown
in fig. \ref{fig:w_vs_T}. The plus symbols are the
50 percent probabilities of an avalanche propagating indefinitely,
below the line, they will die out. The solid line is a best fit
for the constant $c$ in $w_0 = c \exp(1/T_0)$. Given the
simplicity of the estimate, the agreement is excellent. 

Local one dimensional models of avalanches with randomness will often not
show a transition to propagation because there is a finite probability
that at some time it will encounter conditions causing extinction. The
difference here is that the temperature field widens with increasing
time. So if a site adjacent to the ergodic region fails to tunnel,
in the mean tunnel time, the temperature field will take longer to die
away with increasing $w$, and therefore the probability of extinction
rapidly goes to zero as the width increases, again for $\nu =0$. 

Therefore for a  {\em zero temperature} magnetic system with no spin damping, we expect 
that an initial disturbance will likely propagate if its size is above
some threshold value, causing a transition between a static configuration
and one at some effective {\em finite temperature}. 
Of course, if there is any coupling to degrees of freedom other than spin, so
that $\gamma > 0$,
the motion will die out and the spins will stop moving. 
In such a case,
the avalanched spins will lose their energy and cool down to
zero temperature. The system is locally being annealed at a finite rate.

One might expect that the inclusion of more realistic non-relaxational dynamics 
would not influence critical properties, because the ergodic region has a
finite length scale for $\gamma > 0$.  If $\gamma$ is small, one would expect
these effects to shrink the critical regime. 
However there is also a possibility that the critical
behavior would be altered, and this deserves closer scrutiny. 
Work on spin systems, with a conserved
order parameter (e.g.  the  Heisenberg model without disorder) 
has shown that precessional motion is relevant to the dynamic critical behavior~\cite{MaMazenko} in
equilibrium.
Also, it is interesting to note that the
few experiments available on hysteresis criticality do not seem to agree
quantitatively with Random Field Ising Model calculations based upon relaxational dynamics~\cite{BergerEtAl}.

We now consider the issue of RPM. For the proof
of it to be valid, a no passing rule must be satisfied~\cite{Sethna}
related to earlier work on charge density waves~\cite{Middleton}.
If there are two configurations with the spins in the first all more negative
than in the second, then continued evolution under the same field will not
guarantee that the second set of spins will remain more negative, as is required
by this no passing rule. This is because the motion of the spins is effectively
thermal and fluctuations can flip a spin in the second system above
the value of the first system. This is very different than the relaxational
dynamics needed to give RPM, where this can never happen. On the other hand,
over a large enough scale, it may be unlikely that a coarse grained variable
will violate RPM but nevertheless it is possible for this to occur.

For the simplified model described above eqn. \ref{eq:DiffSource}, 
finite damping $\nu > 0$ can be analyzed.
In this case it is quite similar
to heat balance models for explosive crystallization~\cite{Gold}, which
show many interesting properties~\cite{Weeks,Provatas}. 
Instead of the temperature field at the front
widening indefinitely, it should be of finite width, leading to a finite
probability, per unit time, of the avalanche dying out.  Thus one expects 
that in one dimension, propagation will always terminate eventually.
In higher
dimensions, because the surface area of the front is increasing, we do
expect to see infinite sized avalanches.

Their are many interesting further questions to investigate. If the total
perpendicular angular momentum is only weakly broken by interactions, 
then what effect does this have on the dynamics? 
How does precessional
motion effect the size of the critical region of avalanche dynamics?
What is the effect of including disorder on the simplified model, eqn. \ref{eq:DiffSource}?
And last, can these considerations be extended to better understand
avalanches in granular media?

In conclusion, we have shown that avalanches and hysteretic behavior in
spin systems are strongly influenced by precession and the strength of
damping, where small damping makes a system more prone to large avalanches. 
However even with no damping term, coupling to spin waves can lead to
the termination of avalanches.
For finite but
small damping, the growth front becomes spread out over a large region,
for which the spins inside can be described, for short times, by an ergodic
system at high temperature and, for longer times, slowly anneal to a low
temperature state.

J.M.D. thanks A.P. Young and Trieu Mai for useful discussions.


\begin{thebibliography}{}
\bibitem{footnote:invariant} In practice, disorder and dipole interactions will break this, albeit
weakly in some cases.
\bibitem{footnote:negativeT} If the energy were positive this would correspond to a
negative temperature~\cite{Purcell} where correlations should become
antiferromagnetic. We have searched for this interesting regime but have
not found any systems displaying this behavior.
\bibitem{footnote:scaling} There is an additional length scale associated
with the lattice constant. Our analysis assumes that initially, the next spin flip comes from the site
nearest the boundary. That is not always the case, because reducing the lattice
constant sufficiently will also allow other sites to avalanche. In this regime
the correct scaling is $r(T_0/2)w^3 = {\rm constant}$.
\end{thebibliography}
\end{document}